\documentclass[apj]{emulateapj}

\newcommand{\msun}{\mbox{${\rm M}_{\odot}$}}
\newcommand{\Msun}{\mbox{${\rm M}_{\odot}$}}
\newcommand{\nlim}{\mbox{${N}(V<20.5)$}}
\newcommand{\ncum}{\mbox{${N}(t' <t)$}}
\newcommand{\mlim}{\mbox{${M}^{\rm lim}_{\rm cl}$}}
\newcommand{\mcompl}{\mbox{${M}^{\rm 80\%}_{\rm cl}$}}

\newcommand{\tmin}{\mbox{${t}_{\rm min}$}}
\newcommand{\mmax}{\mbox{${M}^{\rm max}_{\rm cl}$}}

\newcommand{\mssp}{\mbox{${M}^{\rm SSP}_{\rm cl}$}}
\newcommand{\missp}{\mbox{${M}^{\rm SSP}_{{\rm cl},i}$}}

\newcommand{\mlssp}{\mbox{${M}_{\rm \lambda}^{\rm SSP}$}}

\newcommand{\mllim}{\mbox{${M}_{\rm \lambda}^{\rm lim}$}}

\newcommand{\mvlim}{\mbox{${M}_{V}^{\rm lim}$}}

\newcommand{\dndt}{\mbox{${\rm d}N/{\rm d}t$}}
\newcommand{\dr}{\mbox{${\rm d}$}}
\newcommand{\Mc}{\mbox{$M_{\rm cl}$}}
\newcommand{\mc}{\mbox{$M_{\rm cl}$}}
\newcommand{\galev}{\mbox{$GALEV$}}
\newcommand{\starburst}{\mbox{$SB99$}}
\newcommand{\tdis}{\mbox{$t_{\rm dis}$}}

\shortauthors{Gieles, Lamers \& Portegies Zwart}
\shorttitle{The Age Distributions of SMC Clusters}

\def\lea{\mathrel{<\kern-1.0em\lower0.9ex\hbox{$\sim$}}}
\def\gea{\mathrel{>\kern-1.0em\lower0.9ex\hbox{$\sim$}}}
\newcommand{\lta}{{\>\rlap{\raise2pt\hbox{$<$}}\lower3pt\hbox{$\sim$}\>}}
\newcommand{\gta}{{\>\rlap{\raise2pt\hbox{$>$}}\lower3pt\hbox{$\sim$}\>}}

\begin{document}
\journalinfo{The Astrophysical Journal}
\submitted{Received 2007 May 9; accepted 2007 June 8}
  \title{ON THE INTERPRETATION OF THE AGE DISTRIBUTION OF STAR CLUSTERS IN THE
  SMALL MAGELLANIC CLOUD}

\author{Mark Gieles}
\affil{Astronomical Institute, Utrecht University, 
  Princetonplein 5, 3584 CC Utrecht, The Netherlands}
\affil{European Southern Observatory, Casilla 19001, Santiago 19, Chile}
\email{mgieles@eso.org}

\author{Henny J.G.L.M. Lamers}
\affil{Astronomical Institute, Utrecht University, 
  Princetonplein 5, 3584 CC Utrecht, The Netherlands}

\author{Simon F. Portegies Zwart}
\affil{ Astronomical Institute `Anton Pannekoek', University of
  Amsterdam, Kruislaan 403, 1098 SJ Amsterdam, The Netherlands, 
  Section Computational Science, University of Amsterdam, Kruislaan
  403, 1098 SJ, Amsterdam, The Netherlands }



\begin{abstract} 
We re-analyze the age distribution (\dndt) of star clusters in the
Small Magellanic Cloud (SMC) using age determinations based on the
Magellanic Cloud Photometric Survey.  For ages younger than
$3\times10^9\,$yr the \dndt\ distribution can be approximated by a
power-law distribution, $\dndt\propto t^{-\beta}$, with
$-\beta=-0.70\pm0.05$  or $-\beta=-0.84\pm0.04$, depending on the
model used to derive the ages.  Predictions for a cluster population
without dissolution limited by a $V$-band detection result in a
power-law \dndt\ distribution with an index of $\sim-0.7$.  This is
because the limiting cluster mass increases with age, due to
evolutionary fading of clusters, reducing the number of observed
clusters at old ages.  When a mass cut well above the limiting cluster
mass is applied, the \dndt\ distribution is flat up to $1\,$Gyr. We
conclude that cluster dissolution is of small importance in shaping
the \dndt\ distribution and incompleteness causes \dndt\ to
decline.  The reason that no (mass independent) infant mortality of
star clusters  around $\sim10-20\,$Myr is found is explained by a
detection bias towards clusters without nebular emission, i.e. cluster
that have survived the infant mortality phase. The reason we find no
evidence for tidal (mass dependent) cluster dissolution in the first
Gyr is explained  by the weak tidal field of the SMC.  Our
results are in sharp contrast to the interpretation of Chandar et
al. (2006),  who interpret the declining \dndt\ distribution as
rapid cluster dissolution. This is due to their erroneous
assumption that the sample is limited by cluster mass, rather than
luminosity.

\end{abstract}



\section{INTRODUCTION}

Star clusters are often assumed to be tracers of the star formation
history of their host galaxy. Recent studies on the star formation
rate in the solar neighborhood have revealed that the majority of
stars form in a clustered environment \citep{2000prpl.conf..151C, 2003ARA&A..41...57L},
while only a few percent of stars in the solar neighborhood are in
clusters. Understanding the process of cluster dissolution is,
therefore, of key importance if one wants to make a meaningful
translation from the observed age distribution of star clusters to a
star formation history of their host galaxy.

Recent theoretical work suggest that a large fraction (50-90\%) of the
star clusters disperse a few Myrs after formation due to the expulsion
of residual gas by the stellar winds and supernovae of massive stars
(e.g. \citealt{1997MNRAS.286..669G, 2003MNRAS.338..673B}).  
If the star formation efficiency is independent of cluster mass (\mc), then
this  ``infant mortality" is also independent of $\mc$
\citep{2006MNRAS.tmp.1213G}.

The observed age
distribution of star clusters in M51 \citep{2005A&A...431..905B}
indicates that $\sim\!70\%$ of the clusters
disperse on a time-scale of $\sim\!10\,$Myr after formation, roughly
independent of \mc. 
\citet{2005ApJ...631L.133F} claim that in the
``Antennae'' galaxies this process removes roughly 90\% of the
clusters each age dex during the first Gyr. This time-scale is far too
long to be explained by the gas expulsion scenario. 

The \mc\ independent dissolution due to gas expulsion is in
sharp contrast to dissolution due to two-body relaxation in a tidal
field (e.g. \citealt{1990ApJ...351..121C,
1997MNRAS.289..898V,2000ApJ...535..759T, 2003MNRAS.340..227B}), or
external perturbations by the disk \citep{1972ApJ...176L..51O,
1997ApJ...474..223G} or giant molecular clouds
\citep{1958ApJ...127...17S, 2006MNRAS.371..793G}, which all have a more destructive effect
on low mass clusters than on high mass clusters.
 
The observed age distributions of star clusters can be used to
 disentangle the two aforementioned disruption processes:
mass independent or mass dependent cluster dissolution.  
If the
disruption time ($\tdis$)  has a power-law dependence on the cluster mass, then the power-law
index can be derived from the slope of the age distribution. This holds for mass
limited and magnitude limited cluster samples, since at old ages
disruption will usually dominate over evolutionary fading
\citep{2003MNRAS.338..717B} (BL03). 
On the other hand, in the case of a {\it \mc\
independent} infant mortality process the age distribution of a {\it
mass limited cluster sample} results in an age distribution of the
form $\dndt\propto t^{-1}$, where the index of $-1$ applies to the case where 90\% of the clusters dissolves each age dex \citep{2005ApJ...631L.133F, 2007AJ....133.1067W}.
 If the sample is magnitude limited,  then the age
distribution will be steeper. This is because fading and
disruption both remove a certain fraction from the \dndt\
distribution.

Recently, \citet{2006ApJ...650L.111C} (CFW06) studied the age
distribution of clusters in the Small Magellanic Cloud (SMC) based on
cluster ages derived by \citet{2005AJ....129.2701R} (RZ05). They claim
that the age distribution is consistent with a power-law with index
$-1$ for $t<3\,$Gyr and that the sample is ``reasonably complete down
to $10^3\,\msun$ over this age range''.  They conclude that their
observations are consistent with infant mortality being at work for
$\sim\!1~$Gyr and independent of \mc.

In this paper we address the interpretation of the \dndt\ distribution of star clusters in the SMC and the pitfall caused by detection incompleteness that can lead to the fallacious conclusion  that mass independent cluster disruption shapes the \dndt\ distribution.

The structure of the paper is as follows. In \S~2 we discuss the 
observations from which the SMC cluster sample was derived and the
mass and age determinations of the clusters. Because the interpretation 
of the age distribution of the cluster sample depends heavily on the (in)completeness
of the sample, we investigate different incompleteness scenarios in \S~\ref{sec3}. In \S~\ref{sec4}
we compare the resulting age distribution to predictions and argue that the shape of the age distribution
it can be explained by evolutionary fading, without involving mass independent
dissolution. In \S~\ref{sec5} we compare our results to other studies and our conclusions are outlined in \S~\ref{sec6}.


\section{OBSERVATIONS}
\label{sec2}

The clusters analyzed by RZ05 were identified by
\citet{2006AJ....131..414H} (HZ06). Photometric data in the $UBVI$ bands from
the Magellanic Clouds Photometric Survey (MCPS;
\citealt{1997AJ....114.1002Z}) were used to find clusters. 
Stellar density images based on the
photometric catalog of SMC stars were constructed by counting the
number of stars with $V<20.5$ in squares of 10\arcsec. We note that
for the photometry all stars in the
catalogue were used, but only for clusters that were identified in the sample
limited to stars with $V<20.5$.  

RZ05 derived ages by comparing the $U-B$, $B-V$ and $V-I$ colors from
clusters using {\it STARBURST99} (hereafter \starburst,
\citealt{1999ApJS..123....3L}) and \galev\ \citep{2002A&A...392....1S,
2003A&A...401.1063A} models.
 We adopt their ages, which have been derived from the photometry with the mean 
foreground extinction correction for the SMC of $E(B-V)=0.09$,
but without a correction for local extinction.
RZ05 tried to include extinction as a free parameter in the age fitting method, but conclude that the scatter in the photometry is too large to improve the age estimates by including extinction as an additional parameter. They take extinction into account in an analytical way\footnote{RZ05 adopted  the mean extinction measured by \citet{2004AJ....127.1531H}  
for stars younger than 10 Myr and older than 1 Gyr. For intermediate ages they 
interpolated the extinction value. RZ05 then corrected the colors for the clusters for which an age estimate is already available, and recalculated the best fit with the predicted clusted models.}, but again conclude that the spread of observed clusters colours around the models does not reduce. Since this artificially imposed extinction as a function of age could introduce systematic trends in the \dndt\ distribution, combined with the conclusion of RZ05 that this method did not reduce the scatter of the cluster colors,  we prefer the first order age estimates of RZ05.

Adopting the ages of the clusters based on their photometry, we derived initial masses of the clusters, 
i.e. corrected for mass loss due to stellar evolution, 
independently using \starburst\, and \galev\, models with
$Z=0.004$. For the \starburst\ models we adopt identical settings as
RZ05, i.e.  standard mass loss, the full isochrone mass interpolation,
a Salpeter initial mass function from $0.1\,\msun$ to $100\,\msun$.

\begin{figure}
\plotone{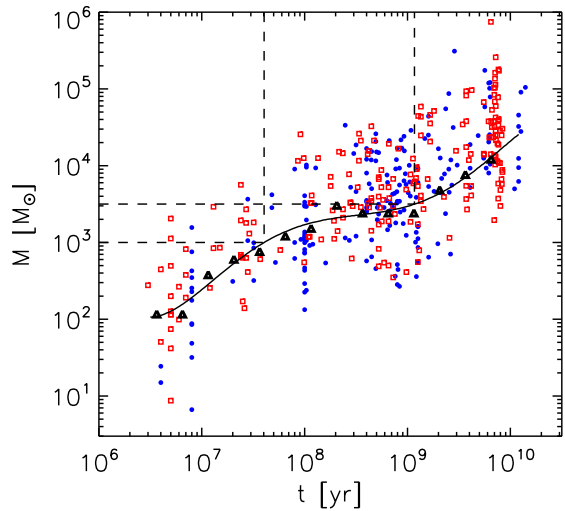}
\caption{Age-mass diagram of 195 clusters in the SMC based on ages derived by 
RZ05 with the \galev\ models (bullets) and the \starburst\ models (squares). 
 The 80\% limit  as a function of age ($\mcompl(t)$) is derived 
from the data (see text for details) and shown as triangles for the \starburst. 
A polynomial fit to $\mcompl(t)$ is shown as a full line. The vertical dashed lines indicate 
the ages where the $\mcompl(t)$ curve intersects with $\log \mc=3$ and $\log \mc=3.5$.
}
\label{fig1}
\end{figure}

In Fig.~\ref{fig1} we show the resulting age-mass diagram for the 195
clusters for which an age estimate is available.  For both samples an
increase in the upper and lower mass boundaries with age can be seen.
The increasing upper mass boundary is due to a size of sample effect, since the
$x$-axis is the logarithm of age.  For a constant cluster formation
rate, more clusters are formed in each dex for increasing $\log t$.
For a power-law cluster initial mass function (CIMF) with index $-2$
the upper boundary is expected to increase as a straight line with
slope +1 \citep{2003AJ....126.1836H, 2006A&A...450..129G}, in reasonable agreement with Fig.~\ref{fig1}.

The sloping lower mass boundary is due to incompleteness, most likely because of evolutionary fading
of clusters with age.
Figure~\ref{fig1} clearly shows that the limiting cluster mass in the sample of RZ05 is not constant,
contrary to the assumption made by CFW06 in their analysis of the age 
distribution. Note that  CFW06 do not present an age-mass diagram in their work to support their assumption, but rely on the assumption that the sample ``is likely to be approximately surface brightness limited, rather than luminosity limited". In \S~\ref{sec3}  we investigate the possible causes for incompleteness and the effect on the \dndt\ distribution in more detail.

Because of the method used by HZ06 to define the cluster sample, the sample of RZ05 is not expected to be 
magnitude limited only (see \S~\ref{sec3}).
In an attempt to quantify the increasing lower \mc\ as a function of age, 
we empirically derive the mass limit above which 80\% of the clusters are
 more massive ($\mcompl(t)$). The limit is chosen at 80\% since we find this to be the point where the mass function (at different ages) turns-over. This suggests the sample is relatively complete above this limit and highly incomplete below it. 
 
At intervals of 0.25 dex in $\log t$ we count the number of clusters in a bin with a width 
of 0.5 dex. 
The masses in the bin are sorted and we count from the highest mass until we have reached 80\% 
of the total number in the bin. Note that this $\mcompl(t)$ is  not the same as an 80\% completeness limit, or the detection limit, as usually derived from artificial cluster experiments, since we here have no information about clusters that did not make it into the sample. However, if we assume that the shape of the completeness curve (i.e. the function that describes the fraction of clusters that is retrieved as a function of mass), is not dependent on age, we can safely assume that  the evolution of $\mcompl$ with age is the same shape as that for the completeness limit.

 The result of $\mcompl(t)$ is shown as triangles in Fig.~\ref{fig1}. 
A polynomial fit  is shown as a full line. 
Note that the curve is located relatively high above the lower limit. If the sample had a ``hard'' detection limit and the cluster IMF
had an index of $-2$ and the cluster dissolution would be mass independent, 
we would expect that at each age the $\mcompl(t)$ line is only a factor 1.25 
higher in mass than the lower limit. Fig.~\ref{fig1} shows that the 80\% limit is at about an order
of magnitude higher mass than the absolute lower limit. 
This implies the data is highly incomplete for low mass clusters.
In the next section we will compare 
the $\mcompl(t)$ curve to different predictions.


\section{Incompleteness: limited by mass, by luminosity or a combination of both?}
\label{sec3}

The interpretation of the empirical age distribution of the SMC clusters
depends strongly on how incompleteness affects the sample. Therefore,
we compare the empirically derived $\mcompl(t)$ curve to what is expected from 
the selection procedure of HZ06 and RZ05.

\subsection{A sample limited by the  number of stars in a cluster}

From the stellar catalogue HZ06 selected stars with $V<20.5$ mag. 
They constructed number density images from which over-densities were detected. 
This  implies that only clusters with enough stars with $V<20.5$ could end up in the catalogue.

We first assume that the number of stars with $V<20.5$ (\nlim) determines the limiting 
cluster mass. We use the evolutionary isochrones of the Padova models for 
$Z=0.004$ \citep{1994A&AS..106..275B,1996A&AS..117..113G,2000A&AS..141..371G} which were 
converted to the $UBVRIJHK$ photometric system by \citet{2002A&A...391..195G}. We assume 
a Salpeter stellar initial mass function (IMF) with lower mass of 0.5 \msun. 
Stars below this mass 
do not reach $V<20.5$ within the Hubble time.
The number 
of stars with $V<20.5$ is counted for cluster models of different ages and masses
and we select only clusters with \nlim\ greater than some fixed number.
We adopted $\nlim = 25$ because the resulting mass-age relation agrees more or less 
with the location of the \mcompl(t)\ line in Fig. 2.

\begin{figure}
\plotone{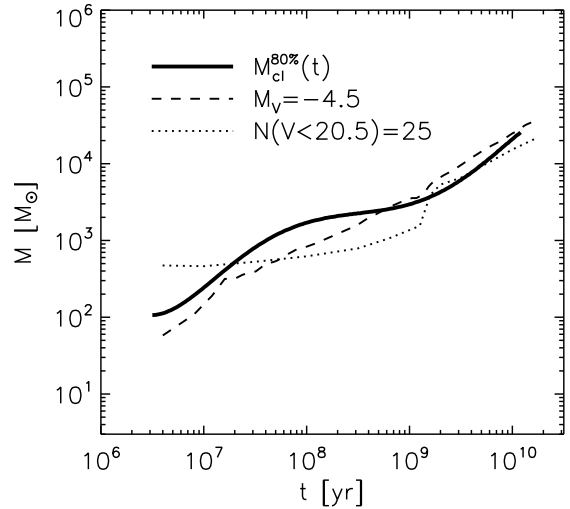}
\caption{The empirically derived $\mcompl(t)$ curve above which 80\% of the clusters
are located  as derived in \S~\ref{sec2} and Fig.~\ref{fig1} (full line). 
Also shown are the two predicted lines for the cluster models with a limiting mass evolution based 
on the assumption of a constant number of stars brighter than $V<20.5$ (dotted line) 
and on the assumption of a limiting cluster magnitude at $M_V < -4.5$ (dashed line).}
\label{fig2}
\end{figure}

In Fig.~\ref{fig2} we show the result as a dotted line. 
The curve is slowly increasing for $\log t\lesssim9$ and then rises. This transition occurs at
the age where the main sequence turn-off drops below $V=20.5$.  Below that age a cluster 
of a given mass has nearly constant \nlim, since the brightness  of stars on the main sequence 
is not varying much and the fraction of stars in \nlim\ that is on the main sequence is high.
 When the turn-off drops below $V=20.5$, \nlim\ consists of RGB and AGB stars mainly.
Because the fraction of RGB and AGB stars in a cluster is small, but is responsible for the majority of the cluster luminosity, the minimum mass of a cluster
with  $\nlim = 25$ steeply increases with age. 

\subsection{A sample limited by the total cluster luminosity}

 After HZ06 selected their clusters from the stellar density image, 
as described in \S~\ref{sec2}, a visual inspection on the 
original images was done. Only the unambiguous clusters were kept. After this,  
they fitted  surface brightness profiles to the remaining sources. 
Only when a reliable surface brightness fit and photometry could be extracted, the 
source was kept in their sample and called a ``star cluster". The total number of sources found from the 
stellar density images was a factor of 4 higher than the final number of star clusters in 
the catalogue (Zaritsky, private communication).  This indicates that accurate photometry
was a strongly limiting factor in the selection of the clusters.
Since this last selection step relies 
on the total luminosity of a cluster we may assume that the sample is mainly limited by
the magnitude of the clusters. To test this assumption,
we predict what is expected for the limiting 
mass as a function of age for a constant magnitude limit.

Star clusters are formed on a time-scale short enough that their
photometric evolution can be well described by Simple Stellar
Population (SSP) models.  From these models it follows that the flux
of a star cluster scales approximately as a power-law with age
($F_\lambda\propto t^{-\zeta}$), with $\zeta$ positive. The limiting
cluster mass as a function of age ($\mlim(t)$) of a sample that is
limited by an absolute magnitude at a certain wavelength ($\mllim$)
can be expressed in the magnitude evolution from an SSP model
($\mlssp(t)$):

\begin{equation}
\log\mlim(t)=\log\missp-0.4\,\left[\mllim-\mlssp(t)\right],
\label{eq:mlim1}
\end{equation}
 where \missp\ is the initial mass of the cluster that is
described by the SSP model. For the \galev\ and \starburst\ version
that we use $\missp=10^6\,\msun$. When taking \missp\ as a constant
$\mlim(t)$ represents an initial mass, i.e. before mass loss due
to stellar evolution. 

Since $\log\missp(t)$
and $\mllim$ from Eq.~\ref{eq:mlim1} are constant and
$0.4\,\mlssp\propto -\log F_\lambda=\zeta\log t$, we can rewrite
Eq.~\ref{eq:mlim1} as

\begin{equation}
\log\mlim(t)=\zeta\log t + C,
\label{eq:mlim2}
\end{equation}
 with $C\equiv\log\missp-0.4\mllim$ (BL03).  
 
 In Fig.~\ref{fig2} we show the
 result of Eq.~\ref{eq:mlim1} as a dashed line  for the \starburst\ models 
 with an adopted limiting magnitude of $\mvlim=-4.5$. This value was chosen because the resulting limiting 
age-mass relation agrees more or less with the location of the 
empirically derived $\mcompl(t)$ limit of Figs.~\ref{fig1} and \ref{fig2}.
Notice that this prediction matches the empirically derived $\mlim(t)$ quite well.
We especially note that this predicted  relation fits the  empirical one
much better than the $\nlim$ curve.

{\it We conclude that the cluster sample of RZ05, used to derive the age distribution
of SMC clusters, is mainly magnitude limited.}

\section{THE AGE DISTRIBUTION OF SMC CLUSTERS}
\label{sec4}

\subsection{The observed age distribution}
\label{subsec41}

The uncertainties in the age determinations of RZ05 are quite large,
mainly due to the photometric uncertainty of faint clusters. We take these
uncertainties into account when constructing the \dndt\ distribution. 

We represent the contribution by each
cluster to the age distribution as an asymmetric Gaussian,
with the lower width ($\sigma_-$) and
the upper width ($\sigma_+$) corresponding to the minimum and maximum age
($t_-$ and $t_+$) derived by RZ05. RZ05 based these latter values on
the 90\% confidence range from their $\chi^2$ results 
of the photometric age determination. This
corresponds to $1.6\sigma$. We assume a Gaussian spread in $\log t$,
since the errors in the age determination are roughly constant in
$\log t$ (mean $\Delta \log t\simeq0.25$).  The definition of $\sigma_\pm$ is then
 $\sigma_\pm\equiv|\log t_\pm-\log t|/1.6$. This results in
an asymmetric Gaussian profile for each cluster, where the left and right side
both have a surface of 0.5, such that the total contribution of each
cluster to the age distribution is 1.  Some ages derived with \starburst\ have $\sigma_-=0$ or
$\sigma_+=0$.  In that case we adopted a minimum uncertainty of
$\sigma_\pm=0.05$.  
When all clusters are added in a large $\log t$ array of equal intervals in $\log t$, 
each value of $\dr N$ (i.e. the number of clusters in that age interval) is
divided by the width of the bin in linear age, to construct the \dndt\
distribution.

\begin{figure*}
\plottwo{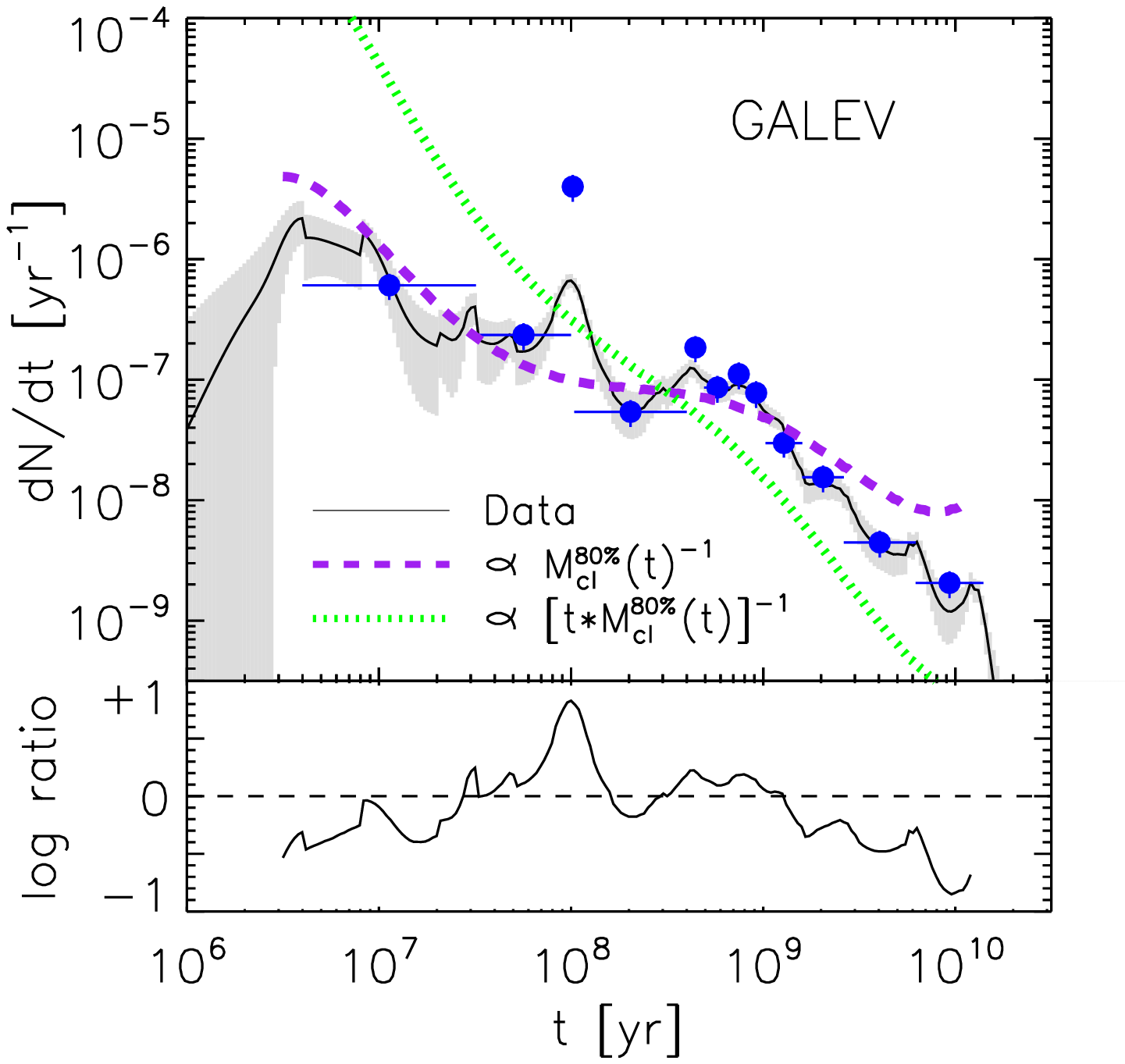}{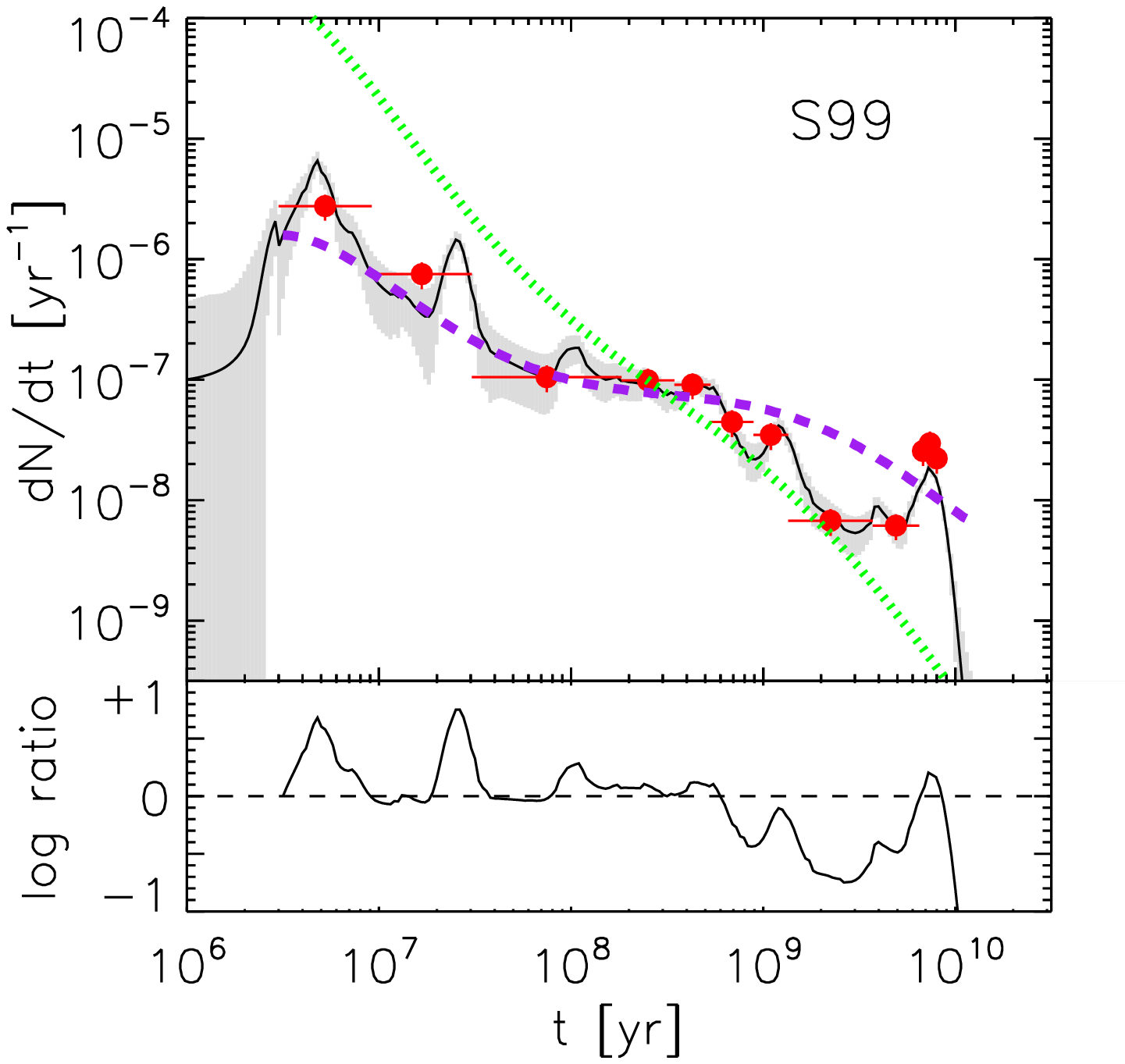}
\caption{Top: The age or \dndt\ distribution from \galev\ (left) and \starburst\ (right) results.
Full line: each cluster was represented by a Gaussian profile with $\sigma_\pm$ corresponding
to the minimum and maximum uncertainty in the age determination. The grey area corresponds to the
$1 \sigma$ uncertainty.
Dots with error bars: the age distribution shown as a histogram where the horizontal lines
indicating the variable width of the age bins, where  each bin contains 17 clusters (see text for details).
Thick dashed line: the predicted \dndt\ distributions for a sample limited by $\mcompl(t)$  of Fig.~\ref{fig1}. The thick dotted line represents a cluster sample limited by $\mcompl(t)$ combined with the mass independent disruption model which was suggested by  CFW06. This distribution declines much more rapidly than the data. 
Bottom panel: the ratio between the observed \dndt\ distribution and the predicted one for the sample 
limited by  $\mcompl(t)$.}
\label{fig3}
\end{figure*}

In the left and right panel of Fig.~\ref{fig3} we present the smoothed
\dndt\ result for the \galev\ and \starburst\ modelling, respectively. 
The full line represents the
full data set. We estimated Poisson errors by counting the number of
clusters in bins of width 0.25 dex, corresponding to the mean uncertainty in the log age values of RZ05. The $1\sigma$ Poisson errors are shown as a grey shaded
region.  Note that we do not normalize the cluster age distribution to the stellar age 
distribution, as was done by RZ05. This to compare our results to the results of CFW06, 
who also construct the unnormalized cluster age distribution.

This representation of the \dndt\ distribution takes into account the increasing
age interval for increasing width of the age intervals. The number of clusters 
per logarithmic age bin actually slightly
increases with $\log t$, which reflects the small increase in number
density that can be seen in Fig.~\ref{fig1}.  Note that $\dndt\propto t^{-1}$ implies
a constant number of clusters in constant $\log t$ bins.

In Fig.~\ref{fig3} we also show a binned \dndt\ histogram as filled
circled with small error bars.  The sizes of the bins are chosen such that
there are $17$ clusters in each bin (following
\citealt{2005ApJ...629..873M}).  This results in a total of 10 bins
over a range of $\sim4.5$ age dex.  The obtained average bin width is
then 0.35 dex, which is larger than the mean error in the age
determination making it suitable for a fit where we only take into account the uncertainties
in the direction of the $y$-axis. 
The first bin starts at the age of the youngest
cluster and the last bin ends at the age of the oldest cluster. This
way of binning the data results in a \dndt\ distribution very similar
to the Gaussian smoothed version (Fig.~\ref{fig3}).

We fit a straight line to $\log(\dndt)$ vs. $\log t$ for $t<3\,$Gyr using a 
$\chi^2$-error statistics minimization with bin weights ($W_i$) depending on the 
standard deviation 
of each bin ($\sigma_i$) as $W_i=1/\sigma_i^2$, which for this representation is 
$W_i=\ln^2(10)\,N_i$. We find a slope, i.e. the power-law index,  of
$-0.70\pm0.05$ if we use the \galev\ results and $-0.84\pm0.04$ if we
use the \starburst\ result. The mean index is thus $0.77\pm0.07$.\footnote{
If we ignore the bin weights in the fit, we find uncertainties on the indices 
of 0.20 and 0.08 
for the fits on \galev\ and \starburst, respectively. This could be the 
explanation for  the 
large uncertainty of 0.15 found by CFW06 in their index of 0.85.}

{\it We conclude that the observed age distribution in the age range of 
$7 \lesssim \log t \lesssim 9.5$ can be approximated by a  power-law with index $-0.77\pm0.07$.}

\subsection{The predicted age distribution}
\label{subsec42}

Assume that a cluster population formed at a constant rate and with a
 power-law CIMF with index $-\alpha$. Then, if there is {\it no dissolution}, 
 the age distribution of all
 clusters can be acquired by integrating over all masses from
 $\mlim$ (from Eqs.~\ref{eq:mlim1} or \ref{eq:mlim2}) to $\mmax$
 (BL03):

\begin{eqnarray}
\dndt&=&\int_{\mlim}^{\mmax} S\Mc^{-\alpha}\dr \Mc\nonumber\\
     &=&\frac{S}{1-\alpha}\left[(\mmax)^{1-\alpha}-(\mlim)^{1-\alpha}\right]\nonumber\\
       &\propto&1/\mlim
\label{eq:dndt}
\end{eqnarray}
 where $S$ describes the cluster formation rate. In the last steps we have used $\alpha=2$
and $\mmax >> \mlim$.
If the sample is magnitude limited, as we have shown in \S~\ref{sec3}, with $\mlim \propto t^{\zeta}$
then the age distribution scales with $t$ as $\dndt \propto t^{-\zeta}$. So we expect a slope $-\zeta$ in the logarithmic representation of the \dndt\ distribution (Fig.~\ref{fig3}).
On the other hand, for a  sample that is  mass limited (corresponding to 
\mlim\ being constant) without dissolution the age distribution would be flat.

 The value of $\zeta$ can be determined by
approximating $\log \mssp(t)$ vs. $\log t$ following from SSP models
by a straight line.  For ages smaller than $3\,$Gyr, the value of
$\zeta$ derived from the \galev\ models are [1.02, 0.89, 0.72, 0.63,
0.58] for $U,B,V,R$ and $I$, respectively. For the \starburst\ models
we find [0.93, 0.78, 0.66, 0.60, 0.54] for the same filters,
respectively. 
So, for a cluster sample formed with constant formation rate, not affected by
disruption and limited by a detection in the $V$ band, the predicted age distribution
is a power-law with index $\sim-0.7$, in agreement with the observations (\S~\ref{subsec41}).

Using the empirically derived $\mlim(t)$  from Figs.~\ref{fig1} and \ref{fig2}, we predict the \dndt\ distribution for a cluster sample that is not affected by any disruption and thus declines as $[\mcompl(t)]^{-1}$ (thick dashed lines in Fig.~\ref{fig3}). The prediction and the observed data are normalized at $\log t =8.5$. The prediction describes the overall decline and the details in the shape of the observed \dndt\ distribution very well. {\it This suggests that there are no signs for cluster disruption present in this data set and the declining \dndt\ distribution can be explained by incompleteness only.}

 To illustrate the combined effect of disruption and incompleteness, we also show the \dndt\ distribution for a sample where we apply the 90\% reduction of clusters each age dex, as suggested by CFW06 (thick dotted line), also normalized to the data at $\log t =8.5$. We stress that the mass independent disruption model of CFW06 works in addition to incompleteness, such that both make the \dndt\ distribution steeper. Therefore, a fallacious assumption that a cluster sample is mass limited, while it is in reality luminosity limited, will lead to a confusion between disruption and incompleteness.

The nice agreement between the observed \dndt\ distribution and the $[\mcompl(t)]^{-1}$ curve implies that {\it the
observed age distribution of the SMC clusters with ages $\log t/{\rm yr} \lesssim 9.5$
can be explained by evolutionary fading only, without the need for mass independent dissolution.}  The fact that there is no evidence for infant mortality at young ages ($\lesssim30\,$Myr) is most likely due to a bias of clusters without nebular emission (HZ06), which typically have such young ages. We expect that the effect of gas removal must destroy young (few Myrs) clusters in the SMC, but we conclude that the sample under discussion here is not suitable to investigate this phenomenon in detail. In a future study we will investigate the infant mortality rate of young star clusters in the SMC using different techniques \citep{gieles07}.

Also tidal evaporation does not affect the \dndt\ distribution for SMC clusters
with ages $\lesssim 3\,$Gyr. This agrees with the long survival times
($\gtrsim6\,$Gyr for masses $\ge10^4\,\msun$) predicted by $N$-body
simulations of clusters dissolving in weak tidal fields (e.g. \citealt{2003MNRAS.340..227B}).

\subsection{The age distribution of a mass limited sub-sample}
\label{subsec43}
The results of CFW06 are based on the assumption that the RZ05 sample
of clusters is mass limited. To show the effect of a mass limit
on the age distribution we compare the cumulative age distribution
of the clusters with and without mass limit to the predictions.
Such a cumulative distribution is preferred over constructing a histogram when only a small number of clusters is available, as is the case when applying a (relatively high) mass cut (see Fig.~\ref{fig1}).

\begin{figure*}
\plotone{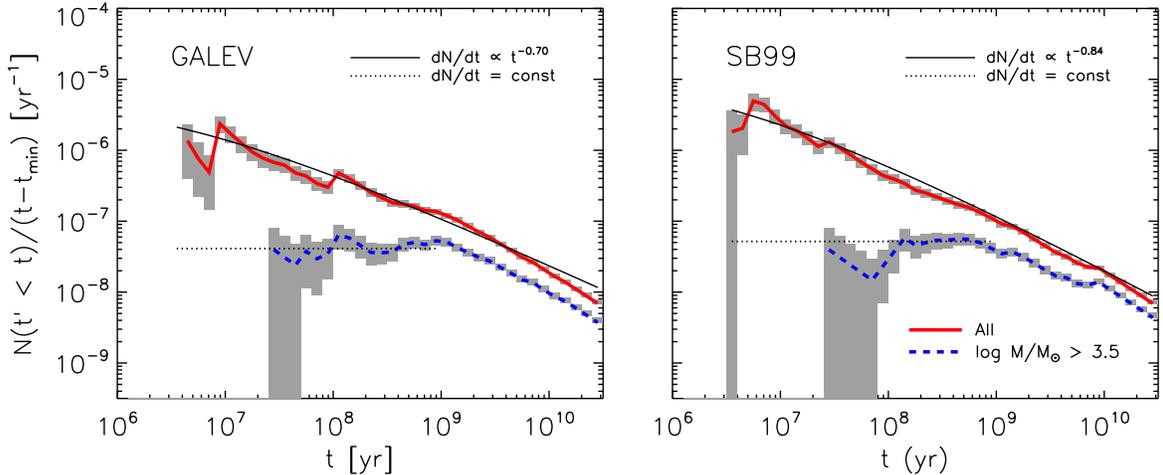}
\caption{The cumulative age distribution (\ncum) normalized to $(t-\tmin)$ for the \galev\ (left) and \starburst\ (right) results. The total sample is shown with thick full lines, with the results of the power-law fits overplotted as thin lines. Eq.~\ref{eq:cum1} was used to relate the power-law function to \ncum. The resulting distributions after a mass cut at $\log M=3.5$ is applied are shown as thick dashed lines. The predictions for a flat \dndt\ distribution, up to the age for which the sample is expected not to be affected by incompleteness ($\sim1\,$Gyr, see Fig.~\ref{fig1}), are also flat in this representation (see text and Eq.~\ref{eq:cum1}  for details) and are shown as dotted lines.}
\label{fig4}
\end{figure*}

Fig.~\ref{fig4} shows the cumulative age distributions divided by the age range from the minimum age in the sample ($\tmin$): $\ncum/(t-\tmin)$. This representation has the advantage that it resembles a \dndt\ distribution. The full SMC cluster sample is shown as thick full lines and the sub-sample of clusters with masses above $3\times10^3$ \msun\  is shown as thick dashed lines, using the 
masses and ages derived by using the \galev\ models (left panel) and the \starburst\
models (right panel). The grey areas of each distribution indicates the $1\sigma$ 
uncertainty due to Poisson statistics. Note that the sub-sample with the mass cut is flat up to $\sim1\,$Gyr, while the full sample is declining. Signs of flattening are also present in the \dndt\ histogram of CFW06 (their Fig.~1) for a sub-sample with $M>10^3\,\msun$. The authors also note this flattening and argue that ``the mass-limited sample is somewhat shallower than the one constructed from the entire sample, since there are relatively few clusters more massive than $10^3\,\msun$ at very young ages." While this is indeed the correct explanation for why the \dndt\ distribution of a mass-limited sub-sample is flatter, it should not be the case in their suggested scenario. In their scenario, the number of clusters in $\log t$ bins of equal width should be constant. Using the \galev\ results, we find 36 clusters with masses above $3\times10^3\,\msun$, and ages between $10^{8.75}\,$yr and $10^{9.25}\,$yr, while there are 0 clusters above this mass limit with ages between $10^{6.75}\,$yr and $10^{7.25}\,$yr. According to the model of CFW06 these numbers should be the same. In addition, we show in Fig.~\ref{fig1} that a mass cut at $10^3\,\msun$ is not high enough to be safely above the increasing limiting mass up to ¼\,Gyr. A mass cut at $3\,10^3\,\msun$ would be a safer choice. Therefore, we will consider this limit.

We compare these cumulative distributions with those predicted for the power-law fits to \dndt. 
If $\dndt=C\,t^{-\eta}$, then the \ncum\ depends on $t$ as

\begin{equation}
\ncum =  \frac{C}{1-\eta} \left[ t^{1-\eta}-t^{1-\eta}_{\rm min} \right ] , 
~~{\rm if}~~\eta\ne1.
 \label{eq:cum1}
\end{equation}
We adopt $\tmin=3\,$Myr, since this is the youngest age present in the SSP models.

 In \S~\ref{subsec41} we have shown that for the full sample $\eta=0.70$ and  
$\eta=0.84$ for \galev\ and \starburst, respectively. 
The resulting predicted cumulative
distribution (Eq.~\ref{eq:cum1}), again normalized to $(t-\tmin$), is shown by the thin full line. Notice that it agrees with the
empirical distribution within about  $2\sigma$. For a sample that is mass limited 
above the detection limit and does not loose clusters by dissolution, i.e. $\eta=0$,
the expected  distribution will be flat in the representation of Fig.~\ref{fig4}.
The thin dotted lines 
in Fig. \ref{fig4} show the predicted distributions. Again, these agree with the
observed cumulative distributions up to an age of $\sim1\,$Gyr, corresponding to the maximum age for which the sample of cluster with $\log M=3.5$ is not affected by the $\mcompl(t)$ line (Fig.~\ref{fig1}). Beyond this age the distribution declines in a similar way as the full distribution, which is because even for the relatively high mass cut, the sample is affected by incompleteness in the same way as the full sample.


\section{COMPARISON TO OTHER STUDIES}
\label{sec5}

We first compare our results to other studies on the SMC star
clusters. This is particularly interesting since there appears to be a
large variation between the different data sets and their
interpretations.

One of the first detailed analyses of the age distribution was done by
\citet{1987PASP...99..724H}. The age distribution presented in that
work looks quite different from the one shown here and in RZ05 and
CFW06. It is flat up to almost 1 Gyr and then has a steep drop. This
is quite similar to what is found in a recent study by
\citet{2006A&A...452..179C} (their Fig.~10) who also present an age distribution for
SMC clusters that is nearly flat the first
$\sim10^8\,$yr. \citet{1987PASP...99..724H} detects stars down to
$B\simeq22-23$ and \citet{2006A&A...452..179C} even down to
$V\simeq24$. Both studies derive their ages with isochrone fitting
techniques. This makes it possible to determine the age even if only a
few stars are available. Assuming that at young ages 
($\lesssim$few 100$\,$Myrs) all clusters are detected, a flat distribution is expected
until the age that the fading or disruption starts to remove clusters
from the sample. Using the sample of \citet{1987PASP...99..724H} 
BL03 argued that the dissolution time of clusters in the SMC is very long and of the 
order of 8 Gyr for a cluster of $10^4$ \Msun. This agrees with the fact that we
do not find evidence for dissolution up to ages of about 3 Gyr from the RZ05 sample. We note that  \citet{2006A&A...452..179C} present 2 age distributions. The one we refer to here is for clusters which were classified C (=genuine star clusters) by \citet{2000AJ....119.1214B}. In their Fig.~7 \citet{2006A&A...452..179C} also present an approximately flat age distribution, defined as $\dr N/\dr \log t$, which corresponds to $\dndt\propto t^{-1}$. However, we note that this sample includes all large OB associations, with sizes much larger ($\gtrsim20\,$pc) than typical star clusters ($\sim5\,$pc).

HZ06 and RZ05 consider only stars with $V<20.5$
for the selection of their clusters. This cut in the stellar sample is
$\sim2-3$ magnitudes brighter than the one on the stellar sample
mentioned before. In addition, the broad band photometry of RZ05 was
based on \citet{1962AJ.....67..471K} and EFF
\citep{1987ApJ...323...54E} profile fits to the surface brightness
profile of clusters. More stars are needed for a good profile fit to the surface 
brightness profile than for an isochrone fit, which results in less  clusters in 
the sample of RZ05  than in the samples of \citet{1987ApJ...323...54E} and \citet{2006A&A...452..179C}.
This idea is supported by the larger number of clusters found 
in the deeper studies of \citet{1987PASP...99..724H} and \citet{2006A&A...452..179C},
 viz. 327 and 311, respectively. This supports our conclusion that incompleteness effects are more important in the sample of RZ05.
 
Comparing our results to those of CFW06, based on the same cluster sample of RZ05,
we note that our interpretation of the age distribution is drastically different.
CFW06 assumed that the cluster sample of RZ05 is mass limited. They fit the age
distribution of the clusters with a power-law of index $-0.85\pm0.15$, and argue that this is consistent within $1\sigma$ with
 mass independent dissolution in which the number of clusters decreases 
by a factor 10 per age-dex, up to an age of about 3 Gyr. We have shown 
that the index of the age distribution is  $-0.77\pm0.07$ (\S~\ref{subsec41}), i.e. consistent with $-1$ only within $3.3\sigma$.
Moreover (and more importantly), we have shown 
in \S~2 that the cluster sample is not mass limited, as is already clear 
from the distribution of the clusters in the age-mass diagram of Fig. \ref{fig1}.
We have given arguments that the cluster sample is approximately magnitude limited.
This follows both from the way in which RZ05 have defined their cluster sample
and from the observed $80\%$ limit of the stars in Fig. \ref{fig1}. This limit
has a very similar slope as that expected for a sample that is limited by the visual
magnitude (Fig.~\ref{fig2}). The observed index of $-0.77\pm0.07$ is very similar to the
one predicted for a sample is magnitude limited in the $V$-band (0.72 for \galev\ 
models and 0.66 for \starburst\ models) if the number of clusters decreases only 
by evolutionary fading below the magnitude limit. 
 We thereby
rule out the need for the infant mortality scenario that proceeds up to
 3 Gyrs as interpreted by CFW06. 
 
 From theoretical arguments
it is expected that infant mortality works on a much shorter
time-scale ($\sim10-20\,$Myr), due to the removal of gas that has not
been used to form stars
(e.g. \citealt{1997MNRAS.286..669G,2003MNRAS.338..673B}). 
The lack of a strong bump in the age distribution of the SMC clusters at young ages ($\lesssim30\,$Myr), as was found for the clusters in M51 \citep{2005A&A...431..905B} and the Antennae galaxies \citep{2005ApJ...631L.133F},
is probably because 
the SMC sample of HZ06 has a  bias towards clusters without 
strong nebular emission.

\section{CONCLUSIONS}
\label{sec6}
We conclude that the observed age distribution of the RZ05 sample of the SMC clusters 
agrees with that predicted for a cluster sample that is magnitude limited
and in which the age distribution in terms of \dndt\ is decreasing with age
by evolutionary fading of the clusters. There is no need to invoke  a
mass independent infant mortality that extends to 3 Gyr and destroys 
90\% of the clusters every age dex, as suggested by CFW06. In fact,
for a magnitude limited cluster sample with an extended independent mass infant mortality, the
age distribution would decrease much steeper with age than observed.

Interestingly, our result also explains the nearly
flat age distribution of the {\it field stars} by RZ05, because not
many clusters are dissolved between $10^7$ and $10^9$ yrs, i.e. after the infant 
mortality phase.  This is in contrast to the interpretation by CFW06 of a very long
mass independent disruption phase, because that model predicts a rapid increase 
in the age distribution of the field stars due to the continuing contribution
of stars from dissolving clusters to the field. Such an increasing
age distribution of the field stars is not observed.

 In this paper we demonstrate that the interpretation of the
cluster age distribution is very sensitive to incompleteness effects
as a function of cluster age and to how these are taken into
account. For extra-galactic (slightly resolved) star clusters various
tests methods have been developed to quantify (in)completeness and its
dependence on cluster luminosity and radius (see
e.g. \citealt{2005A&A...431..905B, 2007A&A...464..495M}).  For nearby
cluster samples such as the one of the SMC discussed here, it should
be possible to construct completeness curves as a function of
structural parameters of the clusters as well. By simulating star
clusters of different ages and masses, thereby taking into account the
luminosity evolution of the individual stars, and adding them to the
catalogues one can retrieve the artificial clusters from the catalogue
using the exact same cluster selection procedure as used for the
data. This procedure is beyond the scope of this paper, since we
mainly want to demonstrate that there is a degeneracy between
incompleteness effects and the mass independent disruption model of
CFW06. 

An important consequence of our result is that it confirms that
cluster life-times are strongly dependent on the environment in which
they evolve, which was already suggested by
\citet{1987PASP...99..724H} based on the difference in the age
distribution for clusters in the SMC and the solar neighbourhood (see
also \citealt{2005A&A...429..173L} for a comparison of cluster
life-times in four galaxies). In contrast to this,
\citet{2007AJ....133.1067W} use the result of CFW06 to show that the
cluster \dndt\ distributions in the Antennae galaxies and in the SMC
are very similar. Although it would simplify things if all clusters
evolve similarly the first Gyr, it is rather counter-intuitive since
then the evolution would not depend at all on the strength of the
tidal field and the number of giant molecular clouds. Our results
support a scenario where the life-time of clusters that have survived
the gas removal phase is determined mainly by environmental factors.


  \acknowledgments{We thank Bruce Elmegreen, Nate Bastian and S{\o}ren
Larsen for discussions and comments that improved the paper. We thank
the referee Dennis Zaritsky for discussion on the cluster
selection procedure and completeness. We are grateful to Evghenii
Gaburov for help and discussions on fitting. This research was
supported in part by the Netherlands Organization for Scientific
Research (NWO grant No. 643.200.503), the Netherlands Advanced School
for Astronomy (NOVA), the Royal Netherlands Academy for Arts and
Sciences (KNAW) and the Leids Kerkhoven-Bosscha fonds (LKBF).}

\bibliographystyle{apj}

\clearpage

\end{document}